\documentclass{PoS}

\usepackage{amsmath}

\newcommand{\be}{\begin{equation}} 
\newcommand{\ee}{\end{equation}} 
\newcommand{\bea}{\begin{eqnarray}} 
\newcommand{\eea}{\end{eqnarray}} 

\newcommand{\V}[1]{\underset{\sim}{#1}}

\title{Charmonium Potentials at Finite Temperature}

\ShortTitle{Charmonium Potentials at Finite Temperature}

\author{\speaker{Chris Allton}\\
       Swansea University\\
       E-mail: \email{c.allton@swan.ac.uk}}

\author{Wynne Evans\\
        Swansea University\\
        E-mail: \email{pyevans@swan.ac.uk}}

\author{Jon-Ivar Skullerud\\
        National University of Ireland Maynooth\\
        E-mail: \email{jonivar@thphys.nuim.ie}}

\abstract{ The charmonium states at non-zero temperature are studied
  on anisotropic lattices with 2 dynamical quark flavours.  Non-local
  operators are used to determine the Nambu-Bethe-Salpeter (NBS)
  wavefunctions via both conventional fitting methods and the Maximum
  Entropy Method.  The interquark potential is determined from the
  solution of the Schrodinger equation, given the NBS wavefunction as
  input following the HAL QCD method. We observe a temperature
  dependent potential which becomes steeper as the temperature
  decreases.  }

\FullConference{The 30th International Symposium on Lattice Field
  Theory - Lattice 2012\\
                June 24-29, 2012\\
                Cairns, Australia}

\begin{document}


\section{Introduction}

The quark-gluon plasma phase of QCD has been studied experimentally at
both RHIC and the LHC. However, a full theoretical understanding of
this phase is still being developed. One quantity which can aid this
understanding is the potential between quarks as a function of
temperature. Due to the widely cited J/$\psi$ suppression
\cite{Matsui:1986dk}, it is natural to consider the interquark
potential of the charmonium system.

There has been theoretical work studying the interquark potential in
quarkonia as a function of temperature from early models
\cite{Karsch:1987pv} to perturbative QCD calculations
\cite{theory-pot}.  Furthermore, there have been some recent
non-perturbative QCD studies (i.e. using lattice simulations) of
interquark potentials which is relevant to the work presented
here. These fall into two categories: (i) non-zero temperature studies
of the {\em static} quark potential \cite{static-pot}; and (ii) {\em
  zero temperature} studies of the potential between quarks with
finite masses \cite{nbs}. This work presents a study of the interquark
potential of charmonium using {\em physical charm quark masses} at
{\em finite temperature} and uses two-flavours of dynamical quarks. A
particular feature of our work is that our lattices are anisotropic
which has the significant advantage in that our correlation function
data is determined at a large number of temporal points.

Our method follows the HAL QCD method \cite{halqcd} of determining the
potential in lattice simulations.\footnote{The original HAL QCD
  programme studied {\em internucleon} rather than {\em interquark}
  potentials from lattice QCD simulations.}  This is based on using
the Nambu-Bethe-Salpeter (NBS) wavefunction as input into the Schr\"odinger
equation and solving for the potential.

In this work we use two methods to determine the wavefunction from our
lattice simulations, conventional exponential fits and the Maximum
Entropy Method. We find that both approaches give qualitatively
similar results. Our main conclusion is that we observe a temperature
dependence in the charmonium potential which is consistent with
expectations, i.e. the potential is steepest for low temperatures,
$T$, and shows signs of flattening at large distances as $T$
increases.

We are in the process of extending our work by using the
``time-dependent'' approach of HAL QCD \cite{time-dep} which determines
the potential directly from the hadron correlation functions \cite{us}.



\section{Schr\"odinger Equation Approach}
\label{sec:nbs}

We follow the HAL QCD method to determine the potential \cite{halqcd}.
We begin by determining the NBS wavefunction of
charmonium $\psi(\V{r})$ from the ($t\rightarrow \infty$ behaviour of
the) correlators of point-split operators,
$J(x,\V{r}) = q(x) \,\Gamma\, U(x,x+\V{r}) \, \overline{q}(x+\V{r})$,
\be
C(\V{r},t) \;=\; \sum_{\V{x}} < J(0;\V{r}) \; J^\dagger(x;\V{r}) >
\;\;\longrightarrow\;\;
|\psi(\V{r})|^2 \;e^{-Et}.
\label{eq:cfn}
\ee
The NBS wavefunction is determined from the matrix element of the
point-split operator,
$\psi(\V{r}) \sim \langle 0 | J(\V{r}) | \text{gnd} \rangle$.
Once the NBS wavefunctions and energies, $E$, are extracted from the
correlators (see section \ref{sec:wfpe}) we use the Schr\"odinger
equation to solve for the potential $V(r)$,

\be
\left( -\frac{\nabla^2}{2\mu} + V(r) \right) \psi(r) = E\; \psi(r).
\label{eq:sch}
\ee
Here, $\mu = \frac{1}{2} m_q \simeq \frac{1}{4} M_{J/\psi}$, is the reduced mass.
We note that this is the opposite approach normally associated with
the Schr\"odinger equation, i.e. the potential is ``reverse
engineered'' from the inputted wavefunction.



\section{Lattice Parameters and Correlators}
\label{sec:lat}

Our lattices are generated with two dynamical flavours of light quarks
using a Wilson-type action with anisotropy of $\xi = a_s/a_\tau = 6$
with $a_s \simeq 0.162$fm and $a_\tau^{-1} \simeq 7.35$GeV
\cite{Oktay:2010tf}.  Table \ref{tab:params} lists the lattice
parameters used.  We note that the range of temperatures, $T=
1/(a_\tau N_\tau)$, is from the confined phase up to $\sim 2T_c$ where
$T_c$ is the deconfining transition. The charm quark is simulated
with the (anisotropic) clover action and its mass is set by matching
the experimental $\eta_c$ mass at zero temperature.

\begin{table}[h]
\begin{center}
\begin{tabular}{ccccr}
\hline
 \multicolumn{1}{c}{$N_s$} & \multicolumn{1}{c}{$N_\tau$} & \multicolumn{1}{c}{$T$(MeV)} &
\multicolumn{1}{c}{$T/T_c$} & \multicolumn{1}{c}{$N_{\rm cfg}$} \\
\hline 
12 & 80 &  90 & 0.42 & 250  \\      
12 & 32 & 230 & 1.05 & 1000 \\ 
12 & 28 & 263 & 1.20 & 1000  \\ 
12 & 24 & 306 & 1.40 &  500  \\ 
12 & 20 & 368 & 1.68 & 1000  \\ 
12 & 16 & 458 & 2.09 & 1000  \\ 
\hline
\end{tabular}
\caption{Lattice parameters used, including spatial and temporal
  dimension, $N_s$ and $N_\tau$.}
\end{center}
\label{tab:params}
\end{table}

\begin{figure}
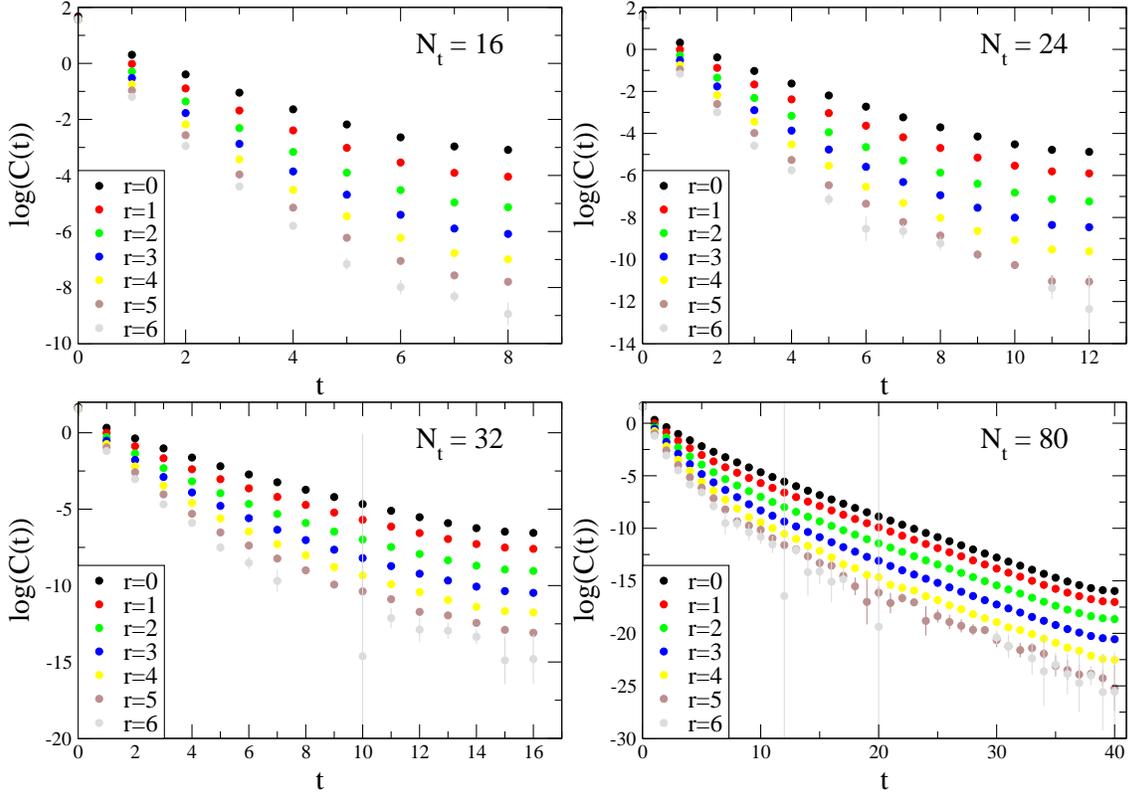

\begin{center}
\includegraphics[width=0.49\textwidth]{logC-vs-t_16.eps}
\includegraphics[width=0.49\textwidth]{logC-vs-t_24.eps}
\includegraphics[width=0.49\textwidth]{logC-vs-t_32.eps}
\includegraphics[width=0.49\textwidth]{logC-vs-t_80.eps}
\caption{Pseudoscalar charmonium correlators of point-split operators
  for various temperatures.}
\label{fig:corfns}
\end{center}
\end{figure}

In Fig.\ref{fig:corfns} we plot the pseudoscalar charmonium
correlation functions for quark separations $r$ at various
temperatures, see table \ref{tab:params}. Only on-axis
separations, $\V{r}$, were studied in this work.

\section{Wavefunctions and Potentials from Exponential Fits}
\label{sec:wfpe}

\begin{figure}
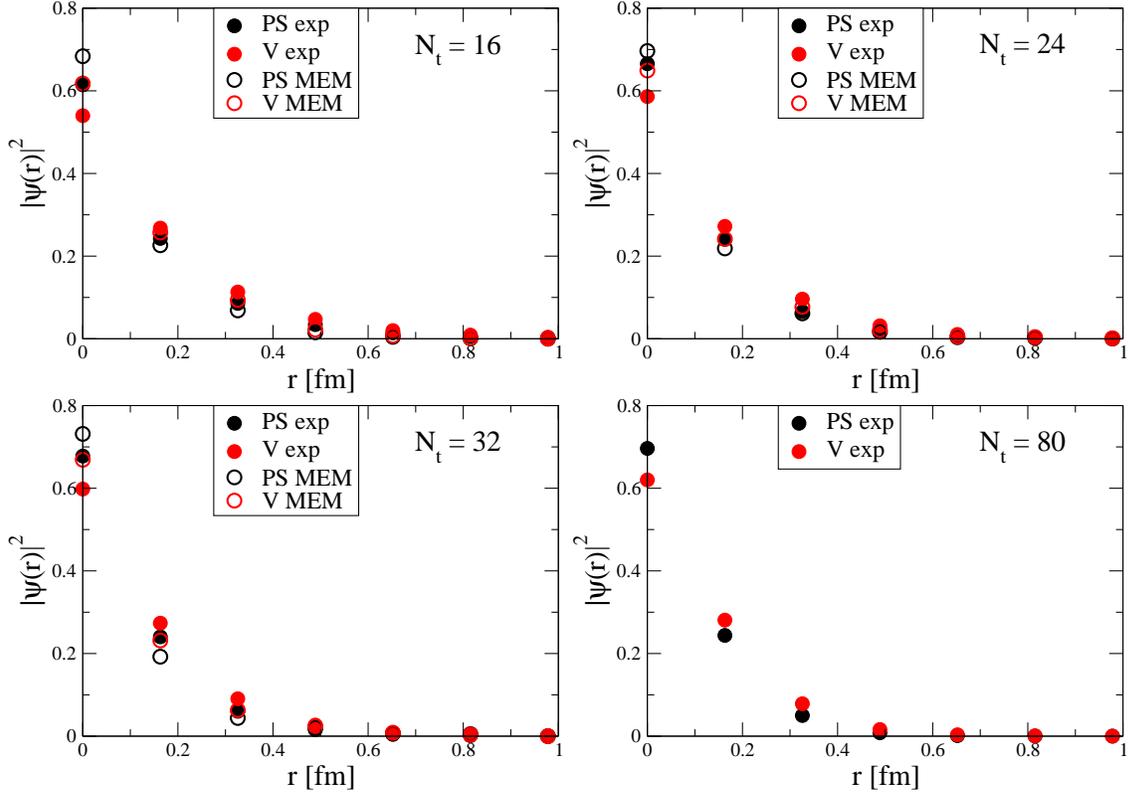

\begin{center}
\includegraphics[width=0.49\textwidth]{wavefn2-vs-r_expmem_16.eps}
\includegraphics[width=0.49\textwidth]{wavefn2-vs-r_expmem_24.eps}
\includegraphics[width=0.49\textwidth]{wavefn2-vs-r_expmem_32.eps}
\includegraphics[width=0.49\textwidth]{wavefn2-vs-r_exp_80.eps}
\caption{Pseudoscalar (PS) and vector (V) normalised wavefunctions,
  $\psi(r)$, for various temperatures.
Both the results from the exponential and MEM fits are shown.
}
\label{fig:wf}
\end{center}
\end{figure}

We extract the NBS wavefunction, $\psi(r)$, discussed in section
\ref{sec:nbs} using a standard exponential fit of the point-split
correlators at large $t$, $C(r,t) \,=\, |\psi(r)|^2 \;e^{-Et}$
(see eq.(\ref{eq:cfn})).  The normalised wavefunctions are plotted in
Fig.\ref{fig:wf} for both the pseudoscalar ($\eta_c$) and vector
($J/\psi$) channels. As can be seen, the expected behaviour for s-wave
ground states is observed with the maximum of $\psi(r)$ at
$r=0$, and $\psi(r) \rightarrow 0$ as $r\rightarrow 0$.

\begin{figure}
\begin{center}
\includegraphics[width=0.49\textwidth]{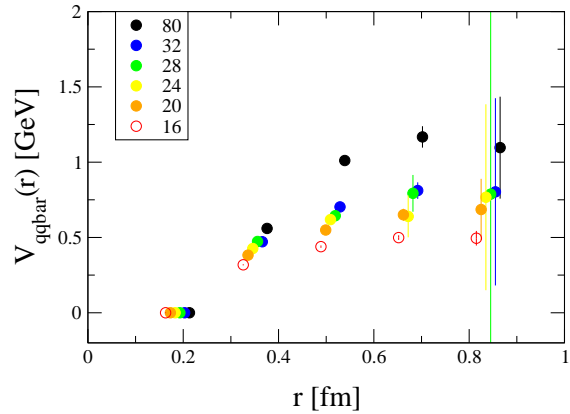}
\caption{The spin-independent potential, $V_{q\overline{q}}$, for
  various temperatures. Statistical errors only are shown, and the
  points are shifted horizontally for clarity. A vertical constant was
  added to each temperature's potential to align the first point,
  i.e. $V_{q\overline{q}}(r/a_s = 1)$ is defined to be zero. The $N_t=16$ points are
  depicted with open symbols due to the concerns about their fits as
  discussed in the text.}
\label{fig:v}
\end{center}
\end{figure}

We use eq.(\ref{eq:sch}) to determine the potential for both the
pseudoscalar and vector channels.  Figure \ref{fig:v} shows the
spin-independent potential, $V_{q \overline{q}}(r)$, defined
\be
V_{q\overline{q}}(r) = \frac{1}{4} [ V_{\rm{PS}}(r) + 3V_V(r) ].
\ee
As can be seen from Fig. \ref{fig:v}, there is evidence of a
temperature dependency in the potential: the potential flattens as the
temperature increases. This is in accord with expectations. Note
however, that the $N_t=16$ correlator fits cannot be made over a large
time range due to the brevity of the lattice in the temporal
direction. For this reason we associate some uncertainty with this
ensemble (and these points are depicted with open symbols in
Fig. \ref{fig:v}).

\section{Wavefunctions from MEM}
\label{sec:mem}

\begin{figure}
\begin{center}
\includegraphics[width=0.49\textwidth]{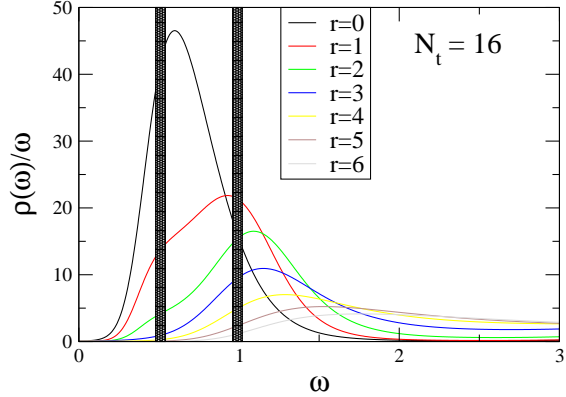}
\caption{The spectral function obtained from MEM for the $N_\tau = 16$
  case for various separations $r$. The vertical bands are placed at
  the position of the ground and excited states.}
\label{fig:rho}
\end{center}
\end{figure}

The Maximum Entropy Method (MEM) has been used by many lattice studies
to extract spectral information from correlators
\cite{Asakawa:2000tr}.  The fundamental equation is
\be
C(r,t) = \int \rho(r,\omega) \; K(t,\omega) \; d\omega,
\ee
where $\rho$ is the spectral function and the lattice kernel is
\be
K(t,\omega) = \frac{\cosh[\omega(t-N_t/2)]}{\sinh[\omega N_\tau/2]}.
\ee
Figure \ref{fig:rho} shows the spectral function obtained via MEM from
the correlators of point-split operators, $C(r,t)$, at the highest
temperature. The vertical bands in Fig. \ref{fig:rho} are placed at
the ground and first excited states' masses as obtained from an MEM
analysis at lower temperatures. We estimate the NBS wavefunction from
the integral of $\rho(\omega)$ over this energy interval.  Figure
\ref{fig:wf} includes the wavefunctions obtained in this way.  As can
be seen there is reasonable agreement between the MEM determination
and the exponential fit approach from section \ref{sec:wfpe}.

Note that the estimate of first excited state's wavefunction from the
second vertical band in Fig. \ref{fig:rho} decreases for small
separations, $r$, as $r \rightarrow 0$, as expected for an excited
state.



\section{Conclusions and Outlook}

There is a significant body of theoretical work studying the
interquark potential in charmonium at non-zero temperature using
model, perturbative and lattice (non-perturbative) approaches.  This
work improves upon earlier lattice simulations by considering quarks
with finite mass, and thus represents a first-principle calculation of
the charmonium potential of QCD at finite temperature. The method we
use is based on the HAL QCD approach which obtains the potential from
correlators of point-split operators \cite{halqcd}.  Our determination
of the potential shows the expected flattening as the temperature
increases.  This work adds to previous charmonium studies performed by
our collaboration with the same lattice parameters
\cite{Aarts:2007pk}.

We've used two different methods of fitting the point-split
correlators: conventional fits to exponentials and the MEM. There is
qualitative agreement between the wavefunctions determined from both.
However, it is clear from Fig \ref{fig:rho} that the (high
temperature) correlators, are not a simple sum of discrete
exponentials, i.e. each spectral feature has a finite width. For this
reason, fitting the correlator to an exponential (as in section
\ref{sec:wfpe}) is an approximation. A better approach would be to
determine the wavefunction from MEM as described in section
\ref{sec:mem}. However, it is necessary to have symmetric
correlators\footnote{i.e. correlators of the {\em same} operator at
  the source and sink} when using MEM to maintain positive
semi-definite spectral weights, $\rho$. Since the operators in
question in this study are point-split and therefore are inherently
noisy, demanding symmetric correlators increases the noise
substantially compared to non-symmetric correlators.

In forthcoming work \cite{us} we will use the HAL QCD ``time
dependent'' approach \cite{time-dep} which allows us to calculate the
potential directly from the correlators, thus circumventing the
problems discussed in both the exponential and MEM fitting procedures.
We also will be studying significantly larger lattices (with a spatial
volume of $32^3$) with 2+1 flavours of dynamical quarks and hope to
extend our work to determine the potential between NRQCD quarks
\cite{Aarts:2012ka}.



\section{Acknowledgements}

We acknowledge the support and infrastructure provided by the Trinity
Centre for High Performance Computing and the IITAC project funded by
the HEA under the Program for Research in Third Level Institutes
(PRTLI) co-funded by the Irish Government and the European Union.  The
work of CA and GA is carried as part of the UKQCD collaboration and
the DiRAC Facility jointly funded by STFC, the Large Facilities
Capital Fund of BIS and Swansea University. WE and CA are supported by
STFC. CRA thanks the Galileo Galilei Institute for Theoretical Physics
for hospitality and the INFN for support during the writing up of this
work. We are very grateful to Sinya Aoki, Balint Jo\'o and Robert
Edwards for useful discussions.





\begin{thebibliography}{99}

\bibitem{Matsui:1986dk}
  T.~Matsui and H.~Satz,
  Phys.\ Lett.\ B {\bf 178} (1986) 416.

\bibitem{Karsch:1987pv}
  F.~Karsch, M.~T.~Mehr and H.~Satz,
  Z.\ Phys.\ C {\bf 37} (1988) 617.

\bibitem{theory-pot}
  Y.~Burnier, M.~Laine and M.~Vepsalainen,
  JHEP {\bf 0801} (2008) 043
  [arXiv:0711.1743 [hep-ph]],
  N.~Brambilla, J.~Ghiglieri, A.~Vairo and P.~Petreczky,
  Phys.\ Rev.\ D {\bf 78} (2008) 014017
  [arXiv:0804.0993 [hep-ph]],
  A.~Dumitru, Y.~Guo, A.~Mocsy and M.~Strickland,
  Phys.\ Rev.\ D {\bf 79} (2009) 054019
  [arXiv:0901.1998 [hep-ph]].

\bibitem{static-pot}
  A.~Rothkopf, T.~Hatsuda and S.~Sasaki,
  PoS LAT {\bf 2009} (2009) 162
  [arXiv:0910.2321 [hep-lat]],
  Phys.\ Rev.\ Lett.\  {\bf 108} (2012) 162001
  [arXiv:1108.1579 [hep-lat]],
  A.~Rothkopf,
  arXiv:1207.5486 [hep-ph],
  Y.~Burnier and A.~Rothkopf,
  Phys.\ Rev.\ D {\bf 86} (2012) 051503
  [arXiv:1208.1899 [hep-ph]].
  A.~Bazavov and P.~Petreczky,
  arXiv:1210.6314 [hep-lat],
  P.~Petreczky, C.~Miao and A.~Mocsy,
  Nucl.\ Phys.\ A {\bf 855} (2011) 125
  [arXiv:1012.4433 [hep-ph]].

\bibitem{nbs}
  T.~Kawanai and S.~Sasaki,
  Phys.\ Rev.\ D {\bf 85} (2012) 091503
  [arXiv:1110.0888 [hep-lat]],
  PoS LATTICE {\bf 2011} (2011) 126
  [arXiv:1111.0256 [hep-lat]],
  Phys.\ Rev.\ Lett.\  {\bf 107} (2011) 091601
  [arXiv:1102.3246 [hep-lat]],
  Y.~Ikeda and H.~Iida,
  PoS LATTICE {\bf 2010} (2010) 143
  [arXiv:1011.2866 [hep-lat]],
  PoS LATTICE {\bf 2011} (2011) 195.



\bibitem{halqcd}
  N.~Ishii, S.~Aoki and T.~Hatsuda,
  Phys.\ Rev.\ Lett.\  {\bf 99} (2007) 022001
  [nucl-th/0611096],
  S.~Aoki, T.~Hatsuda and N.~Ishii,
  Prog.\ Theor.\ Phys.\  {\bf 123} (2010) 89
  [arXiv:0909.5585 [hep-lat]],
  S.~Aoki {\it et al.}  [HAL QCD Collaboration],
  arXiv:1206.5088 [hep-lat].


\bibitem{Oktay:2010tf}
  M.~B.~Oktay and J.~-I.~Skullerud,
  arXiv:1005.1209 [hep-lat].

\bibitem{time-dep}
  N.~Ishii [HAL QCD Collaboration],
  PoS LATTICE {\bf 2011} (2011) 160.

\bibitem{us}
W.~Evans, C.R.~Allton, J.-I.~Skullerud, {\em in preparation}.

\bibitem{Asakawa:2000tr}
  M.~Asakawa, T.~Hatsuda and Y.~Nakahara,
  Prog.\ Part.\ Nucl.\ Phys.\  {\bf 46} (2001) 459
  [hep-lat/0011040].

\bibitem{Aarts:2007pk}
  G.~Aarts, C.~Allton, M.~B.~Oktay, M.~Peardon and J.~-I.~Skullerud,
  Phys.\ Rev.\ D {\bf 76} (2007) 094513
  [arXiv:0705.2198 [hep-lat]].

\bibitem{Aarts:2012ka}
  G.~Aarts, C.~Allton, S.~Kim, M.~P.~Lombardo, M.~B.~Oktay,
  S.~M.~Ryan, D.~K.~Sinclair and J.~-I.~Skullerud,
  arXiv:1210.2903 [hep-lat].

\end{thebibliography}
\end{document}